\documentclass[usenatbib]{mn2e}
\usepackage{natbibmnfix, graphicx, times, amsmath, epsfig, amssymb, amsfonts, textcomp}

\newcommand\lsim{\mathrel{\rlap{\lower4pt\hbox{\hskip1pt$\sim$}}
        \raise1pt\hbox{$<$}}}
\newcommand\gsim{\mathrel{\rlap{\lower4pt\hbox{\hskip1pt$\sim$}}
        \raise1pt\hbox{$>$}}}
\def\myputfigure#1#2#3#4#5%
{\vskip#5pt\makebox[0pt]{\hskip#2in
\includegraphics[width=#3\textwidth]{#1}}\vskip#4pt\hfill}

\pdfoutput=1

\begin{document}

\title[Some comments on the electrodynamics of binary pulsars]{Some comments on the electrodynamics of binary pulsars}
\author[Sobacchi \& Vietri]{Emanuele Sobacchi$^1$\thanks{email: emanuele.sobacchi@sns.it}, Mario Vietri$^1$\thanks{email: mario.vietri@sns.it}  \\
$^1$Scuola Normale Superiore, Piazza dei Cavalieri 7, 56126 Pisa, Italy\\
}

\voffset-.6in

\maketitle

\begin{abstract}
We consider the electrodynamics of in-spiraling binary pulsars, 
showing that there are two distinct ways in which they may emit radiation.
On the one hand, even if the pulsars do not rotate, we show that {\it in
vacuo} orbital rotation generates magnetic quadrupole emission, which, in
the late stages of the binary evolution becomes nearly as effective as 
magnetic dipole emission by a millisecond pulsar. On the other hand, we
show that interactions of the two magnetic fields generate
powerful induction electric fields, which cannot be screened by a suitable 
distribution of charges and currents like they are in isolated pulsars. 
We compute approximate electromotive forces for this case. 
\end{abstract}


\section{Introduction}
\label{sec:intro}

The recent discovery of the binary pulsar J0737-3039A/B \citep{Burgay03, Lyne04}
has highlighted the fact that pulsars may retain their 
magnetic fields for at least a significant fraction of the time it takes
them to in-spiral to their eventual merger. 

This discovery has been used so far to illustrate the extraordinary opportunity 
to study hitherto unaccessible General Relativistic effects, while less 
attention has been paid to the electromagnetic interactions of the two 
magnetospheres (but see \citealt{KS08} also for a discussion of  
magnetospheric effects). 

But even before this discovery, \citet{Vietri96} and \citet{HL01}
have discussed
some of the consequences of assuming that the two inspiralling, merging
{\it pulsars}, ({\it i.e.}, not just neutron stars) may produce observationally 
interesting electromagnetic signals. In particular, \citet{HL01} considered
the magnetospheric interaction of two inspiralling pulsars with very different 
magnetic fields: a fast, but weak-field pulsar, and a slow magnetar. 

In this paper, we wish to take a different point of view, that in which
the two fields do not differ by four to six orders of magnitude like in
\citet{HL01}, but are more evenly matched, though not necessarily of the
same order of magnitude. In particular, we wish to stress first that the
ultimate energy reservoir to be tapped is the binary orbital motion, 
which we do by considering the simplest model possible (two point-like
dipoles orbiting their common center of mass {\it in vacuo}), and pointing
ou that this simple estimate alone leads to an observationally interesting
signal. The reason is that binary pulsars will 
rotate ever faster as a consequence of the orbital decay induced by  
gravitational radiation. If the pulsars manage to retain some signifiant 
fraction of their magnetic fields 
until the moment of merging, their evolution will lead naturally to a  sub-millisecond 
object emitting copious amounts of radiation in the last instants 
before merging, even when we neglect the pulsars' rotation around their spin
axes. We shall first discuss this emission mechanism, which occurs 
{\it in vacuo}. 

On the other hand, under these circumstances there is a natural 
mechanism that will create an induction electric field with a component 
directed along the magnetic field, for an arbitrary orientation of the axes 
in question. This mechanism follows from the well--known inability of the two pulsars to 
synchronize their spin periods with their orbital period, a direct consequence 
of their tiny sizes (in terms of volume, not mass, of course; see \citealt{BC92}). 
Since sufficiently late in 
the binary life the orbital period is much shorter than the spin periods,  we 
may picture the two pulsars as non-rotating; still, the orbital motion swings 
each magnetosphere by the other one, so that any particular part of it is 
compressed on the day-side, and is free to expand on the night-side.  This 
magnetospheric {\it pumping} implies a time variability of the local magnetic 
field, and in turn this leads to an induction electric field. 

This situation differs considerably from the aligned rotator \citep{CKF99}, 
where a judicious choice of a stationary 
charge distribution inside the magnetosphere can short out the component of 
the electric field along the magnetic field; we shall show later that this 
mechanism cannot be effectively screened, as it is in the case of an
isolated, rotating aligned pulsar (IRAP for short). This suggests that large-scale, 
large-amplitude electric fields not necessarily orthogonal to the magnetic field 
do exist in the magnetospheres of binary pulsars. 

It is the aim of this paper to investigate some consequences of this simple 
idea for the (photonic) observability of in-spiraling binary pulsars as GWR 
pushes them closer and closer. In the next Section, we will derive the
energy lost per unit time by two co-rotating magnetic moments {\it in vacuo}, a 
generalization of \citet{Pacini67}'s formula for the isolated magnetic moment.
In Section 3, we consider in some detail the induction effect briefly 
discussed above, still using the {\it in vacuo} case as our model, 
for lack of a fully satisfactory realistic model. We shall discuss how the 
component of the induction electric field along a magnetic field line cannot 
be shorted by arbitrary charges and currents located 
in the magnetosphere, and we shall provide some exact and some approximate 
estimate of the available electromotive forces. The last Section summarizes
our results.

\section{In vacuo}

The starting point of this paper is a computation of the amount of 
energy radiated per unit time by a configuration of two point-like dipoles 
$\vec\mu_1,\vec\mu_2$ rotating around their common center of mass. We assume in 
this that the stars carrying the magnetic moments are not spinning at all 
($\Omega = 0$), and call $\omega \neq 0$ the orbital angular frequency. 

This configuration does not emit magnetic dipole radiation, because
the total magnetic moment $\vec\mu_1+\vec\mu_2$ is a constant, thus 
to this order an observer located at infinity will perceive no 
time-varying field. However, said observer will perceive a 
time--varying field to higher order, due to the fact that the 
magnetic field is stronger when the star with the stronger 
dipole moment is closer, and weaker half an orbital period 
later when it is furthest, and the weaker dipole is closest. 

This effect depends on the orbital radius $r$: the 
modulation of the fields vanishes for $r=0$, and it is clearly 
linear in $r$ for $r \ll $ the observer's distance. Thus 
the radiated power will be quadratic in $r$, showing that we are
dealing with quadrupole magnetic radiation. 

It is obviously possible to obtain the radiated power by means of a
vector-spherical harmonics analysis \citep{Jackson75}. However it is 
also possible to reach the same result via a simpler approach, which
also allows us to introduce the fields we shall use in the next 
Section. 

\citet{Monaghan68} has given exact expressions for the electromagnetic
potentials and fields generated by an arbitrarily moving dipole, i.e., a
particle endowed with both electric ($\vec p$) and magnetic 
($\vec \mu$) dipoles. We just need to specialize to the case 
when, in the particle (actually a star, in our case) reference 
frame, it has only a non--vanishing magnetic dipole moment; in
this case Monaghan shows that, when the pure dipole magnetic
moment is moving with arbitrary speed $c\vec\beta$, it appears 
to have an electric dipole electric moment 
\begin{equation}
\label{eq:onlymu}
\vec p = \vec\beta\wedge\vec\mu\;.
\end{equation}

In this case the electromagnetic potentials are:
\begin{equation}
  \vec A = \left(\frac{\vec\mu\wedge\hat n}{K R^2}+\frac{d}{d t}\frac{\vec\mu\wedge\hat n}{K R c}
  +\frac{d}{dt}\frac{\vec\beta\wedge\vec\mu}{K R c}\right)_{\rm ret}
 \end{equation}
 \begin{equation}
  \phi = \left(\frac{(\vec\beta\wedge\vec\mu)\cdot\hat n}{K R^2}+\frac{d}{dt}
  \frac{(\vec\beta\wedge\vec\mu)\cdot\hat n}{K R c}\right)_{\rm ret}
 \end{equation}
where $K = 1-\hat n\cdot\vec\beta$, $\vec\beta = \vec v/c$ and $\vec\mu$ is the magnetic
dipole of the pulsar. 

\citet{Monaghan68} also gives an especially useful expression for the 
actual fields:
\begin{eqnarray}
\label{eq:Eexact}
 \vec E = \left[
 \frac{3(\vec p\cdot \hat n)\hat n -\vec p}{R^3} +
 \frac{R}{c}\frac{d}{d t}\left(\frac{3(\vec p\cdot\hat n)\hat n -\vec p}{R^3}\right)\right.  \nonumber \\ \left.
 -\frac{d}{dt}\left(\frac{3(\vec p\cdot\hat n)\hat n -\vec p}{c^2 R^2}\frac{dR}{d t}+
 \frac{\vec\mu\wedge\hat n}{K c R^2}\right)\right.  \nonumber \\ \left.
  -\frac{1}{c^2}\frac{d^2}{dt^2} \left(\frac{\hat n\wedge (\hat n\wedge \vec p) -\vec\mu \wedge\hat n}{KR}\right)\right]_{\rm ret} 
\end{eqnarray}
where of course eq. \ref{eq:onlymu} applies. Correspondingly, one 
also obtains:
\begin{eqnarray}
\label{eq:Bexact}
 \vec B = \left[
 \frac{3(\vec \mu\cdot \hat n)\hat n -\vec \mu}{R^3} +
 \frac{R}{c}\frac{d}{d t}\left(\frac{3(\vec \mu\cdot\hat n)\hat n -\vec \mu}{R^3}\right)\right.  \nonumber \\ \left.
 -\frac{d}{dt}\left(\frac{3(\vec \mu\cdot\hat n)\hat n -\vec \mu}{c^2 R^2}\frac{dR}{d t}-
 \frac{\vec p\wedge\hat n}{K c R^2}\right)\right.  \nonumber \\ \left.
  +\frac{1}{c^2}\frac{d^2}{dt^2} \left(\frac{\hat n\wedge (\hat n\wedge \vec\mu) +\vec p\wedge\hat n}{KR}\right)\right]_{\rm ret} 
\end{eqnarray}
The reason why these expressions are useful is that all radiation 
terms are contained in the last term on the rhs of eq. \ref{eq:Eexact} (and \ref{eq:Bexact}):
\begin{equation}
 \vec E_{\rm rad} = -\frac{1}{c^2}\frac{d^2}{d t^2} \left( 
 \frac{\hat n\wedge(\hat n\wedge\vec p) - \vec \mu\wedge\hat n}{KR}
 \right)_{\rm ret}\;.
\end{equation}
Like we said above, we completely neglect the star's rotation, so that $\vec\mu$ 
is a constant and $\vec p = \vec \beta\wedge\vec \mu$. Expanding the above 
expression to lowest order in $\beta$ we find:
\begin{equation}
 \vec E_{\rm rad} \approx \frac{(\hat n\cdot\mu) (\vec{\ddot\beta}\wedge\hat n)}{R c^2}
\end{equation}
We can now specialize to the case of a binary pulsar, each with magnetic moments
$\vec\mu_1$ and $\vec\mu_2$, $\hat n_1 = \hat n_2$, and $\vec\beta_1 = - \vec\beta_2 = 
\vec\beta$ (for equal mass stars, of course)  to obtain
\begin{equation}
 \vec E_{\rm rad} = \frac{(\hat n\cdot(\vec\mu_1-\vec\mu_2)) (\vec{\ddot\beta}\wedge\hat n) }{R c^2}\;,\;
 \vec B_{\rm rad} = \hat n\wedge \vec E_{\rm rad}\;.
\end{equation}
From Poyinting's vector, $c \vec E_{\rm rad}\wedge \vec B_{\rm rad}/4\pi$ we obtain the total radiated 
power as
\begin{equation}
\label{eq:Pint}
 P_{\rm M} = \int d\!\Omega \frac{(\hat n\cdot(\vec\mu_1-\vec\mu_2)^2 (\vec{\ddot\beta}\wedge\hat n)^2}{4\pi c^3}\;.
\end{equation}
We can now, first average over the orbital phase, then carry out the integral over angles 
(see the Appendix) to obtain:
\begin{equation}
\label{eq:Ptotal}
 P_{\rm M} = \frac{\omega^4 \beta^2}{15 c^3}\left(3(\mu_1-\mu_2)_\perp^2 + 4(\mu_1
 -\mu_2)_{\rm z}^2 \right)
\end{equation}
where $\mu_{\rm z}$ the component of the magnetic field along the orbit's normal, 
while $\mu_\perp$ is the component in the plane of the orbit. 

This formula clearly describes the radiation from a magnetic quadrupole, it is
a full factor $\beta^2$ smaller than the dipole term. It has
partially been derived before: \citet{HT75} studied 
the radiation from a pulsar with a (single) magnetic dipole displaced by a fixed
amount from the rotation
axis. In their case, they had dipolar radiation because the components of 
$\vec\mu$ in the orbital plane do not remain constant as a consequence of the 
star rotation. However, the $\mu_{\rm z}$ component produces a quadrupole term
which they computed, and coincides with our result above. 

We now specialize to the case of a binary pulsar whose orbital decay is driven 
by gravitational wave radiation losses, in which case (as we are about to see) 
losses are completely dominated by GW. It can be seen from the equation 
above that electromagnetic losses depend on the star-to-star distance $a$ 
as $a^{-7}$, and are thus strongly peaked around the first moment of physical 
contact; we may thus safely assume that the orbit has already been circularized 
by GWs, and consider only circular orbits. We have 
\begin{equation}
\label{eq:omega}
 \omega^2 = \frac{GM}{a^3}\;\;\;,\;\;\; E = -\frac{1}{2}\frac{G{\cal M}M}{a}
\end{equation}
where the total mass and reduced mass are given, respectively, by 
$M = M_1 + M_2$,  ${\cal M} = M_1 M_2/M$; the GW loss rate is 
\begin{equation}
 \dot E_{\rm GW} = \frac{32}{5} \frac{G^4}{c^5}\frac{M^3{\cal M}^2}{a^5}\;.
\end{equation}
From the above we find
\begin{equation}
 \frac{d a}{d t} = - \frac{2 \dot E_{\rm GW} a^2}{G M {\cal M}}\;.
\end{equation}
Thanks to the equation above, in the limit $M_1=M_2$, we can transform eq. \ref{eq:Ptotal} 
into
\begin{equation}
 \label{eq:Edertotal}
 \frac{d E_{\rm M}}{d a} = \frac{1}{384} \frac{\mu^2 x}{a^4}
\end{equation}
where we have called $E_{\rm M}$ the amount of orbital energy lost via electro-magnetic processes
($P_{\rm M} \equiv dE_{\rm M}/dt$) and 
\begin{equation}
 x \equiv \frac{3(\vec\mu_1-\vec\mu_2)_\perp^2 + 4(\mu_{\rm 1z}-\mu_{\rm 2z})^2}{\mu_1^2+\mu_2^2}
\end{equation}
and $\mu^2 \equiv \mu_1^2+\mu_2^2$. Integrating over $a$ we obtain the total amount of 
radiation emitted by the magnetic quadrupole term:
\begin{equation}
 \label{eq:EMtotal}
 E_{\rm M} = \frac{1}{1152}\frac{\mu^2 x}{a_{\rm min}^3}\;.
\end{equation}
For typical values $\mu = 10^{30}\text{ G cm$^3$}$, $x 
\approx 3$ and $a_{\rm min} = 2\times 10^6\text{ cm}$, we obtain
\begin{equation}
 E_{\rm M} \approx 6.5\times 10^{38} \left(\frac{\mu}{10^{30}\text{ G cm$^3$}}\right)^2\text{ erg}\;.
\end{equation}

This amount of radiation must accompany the final stages of a binary pulsar.
It is small compared to the total amount radiated away by GWs, as we anticipated.
It is strongly concentrated toward the latest moments in the binary existence, 
because $P_{\rm M} \propto a^{-7} \propto t^{-7/4}$. The above estimate assumes 
that fields as large $10^{12} \text{ G}$ exist up to the time of merger. However, even 
a field as low as $10^{10}\text{ G}$ may produce a detectable amount of radiation. 
Thus the above mechanism essentially predicts the existence of radio bursts of roughly
millisecond duration.

Though admittedly the estimate above is a text-book exercise, we have been unable
to find it in the literature. Its importance lies in the illustration of the fact that 
the ultimate energy reservoir is the binary's orbital energy, and in the fact that, even 
under highly idealized assumptions, it leads to a potentially detectable signal.  

Still, we show below that another charge-acceleration mechanism exists, which 
is likely to lead to an estimate of radiated electromagnetic energy dwarfing 
this one.

\section{Induction}

\begin{figure*}
\vspace{+0\baselineskip}
{
\includegraphics[width=0.45\textwidth]{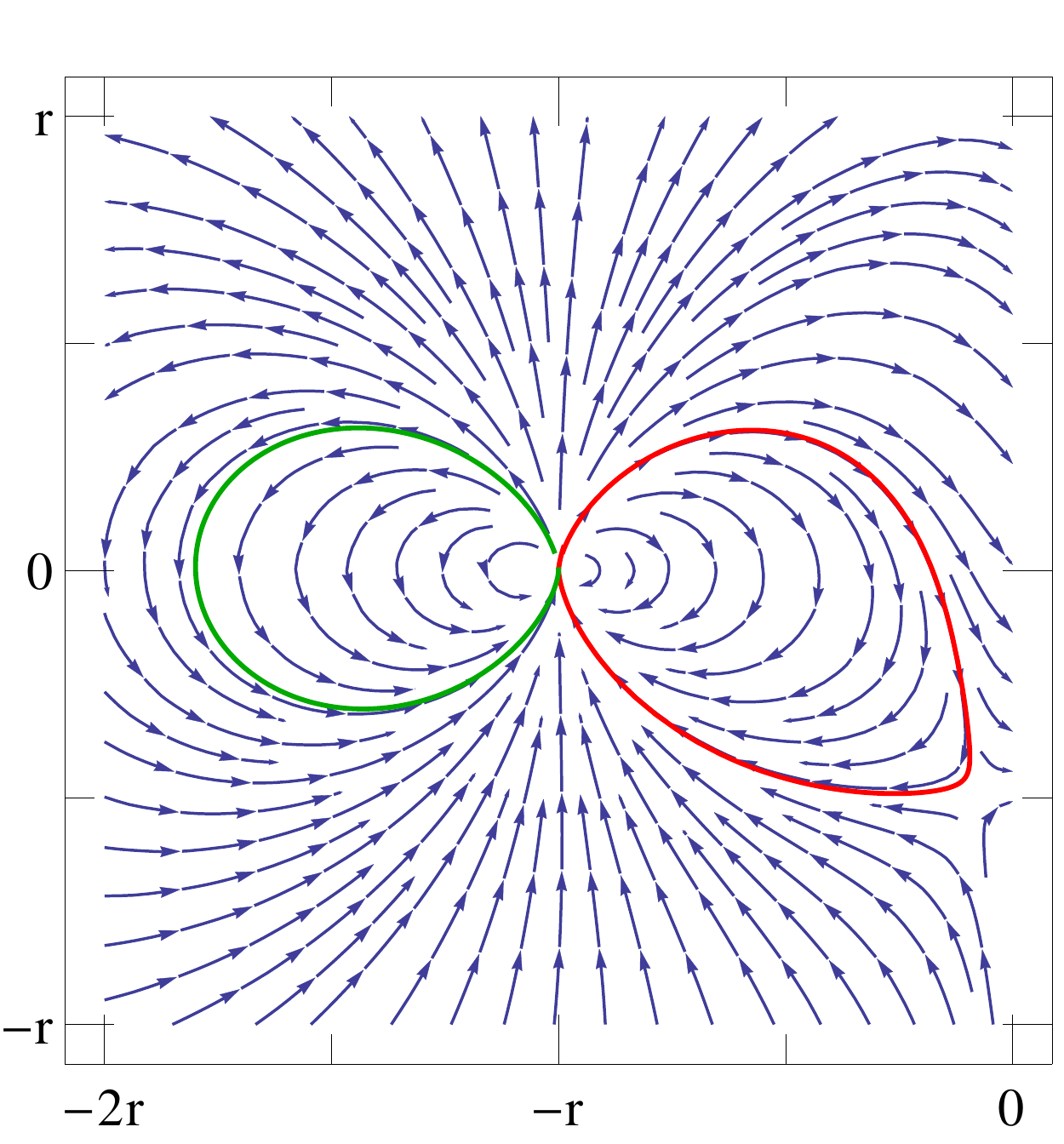} 
\includegraphics[width=0.45\textwidth]{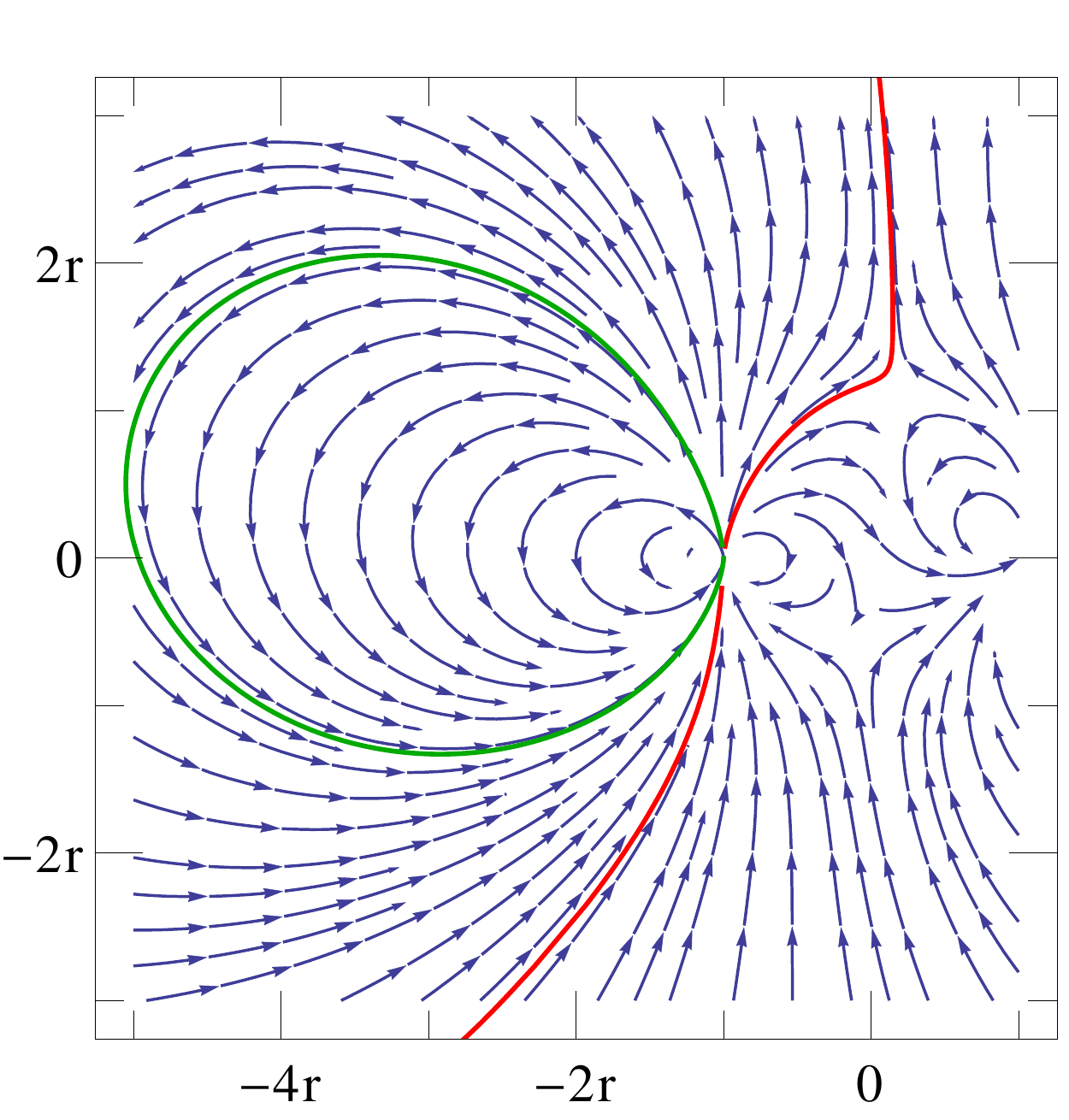} 
}
\caption{Deformation of field lines from dayside (red) to nightside (green) for 
two identical, point-like dipoles in $(-r;0)$ and $(r;0)$ oriented (respectively) at 
90\textdegree and 45\textdegree; the line in the upper panel, if unperturbed, 
is closer to the surface of the star.
\label{fig:field}
}
\vspace{-1\baselineskip}
\end{figure*}

In order to see why induction electric fields will be present when both
stars have a sizable magnetic field, we consider first two stars {\it
in vacuo}, neglecting the stars' rotation but including their orbital motion. 
In this case, the side of the field which is brought by orbital motion closer to the
companion is compressed by the companion's magnetic field, but, 
as theorbital motion leads it away from the companion, it is free to expand again. 
It is this rhythmic compression which generates locally a transient 
magnetic field, hence an induction electric field. 

This periodic effect is illustrated in Fig. \ref{fig:field}, where
the position of a magnetic line has been computed for identical, 
point-like magnetic dipoles. 

This effect exists also in the opposite limit, when the orbital period
is much longer than at least one of the spin periods: again, when a
given portion of the magnetosphere is on the day-side it is compressed
by the companion's field, while it expands as it moves toward the 
night-side. 

There are two special circumstances under which this effect does not 
take place. If the stars could synchronize their orbital and rotational 
motions, thanks for instance to tidal forces, then they would present to 
their companion always the same portion of their magnetic field, which 
would not be rhythmically compressed. However, following the work of \citet{BC92},
we know that neutron stars are too small for 
tidal forces to produce orbit-spin synchronization. 

Also, the induction field vanishes if the two spin axes and the 
orbital angular momentum are parallel, as a consequence of 
the system's reflection symmetry. 

We now consider what happens when we allow for the presence of 
free charges inside the magnetosphere. At this point, one might 
reason as \citet{GJ69}: given the abundance of 
free charges, and the fact that they move freely along magnetic 
field lines, they may redistribute themselves so as to short out 
the component of the electric field along $\vec B$, yielding 
the transversality condition,
\begin{equation}
 \label{eq:transversality}
 \vec E \cdot \vec B = 0\;.
\end{equation}
This hunch is fully borne out by the numerical solution of
\citet{CKF99} in the (stationary!)
case of the isolated, rotating aligned pulsar (IRAP for short). 
In a time-dependent situation, the same spontaneous redistribution
of charges will try once again to enforce the transversality 
condition, eq. \ref{eq:transversality}, and it will succeed
to the extent that we may assume charge carriers to be 
infinitely fast, but we may safely expect this to fail, to some
extent, when account is taken of the finite speed of charge carriers. 

This situation is similar to the penetration of electromagnetic
waves into a good conductor, a textbook problem \citep{Jackson75}: 
so long as the speed of the charge carriers is assumed infinite, there 
can be no penetration of the wave fields inside the conductor, but when
account is taken of either the finite speed of the charge carriers,
or of the existence of a non-zero, albeit small,
resistivity, then the wave fields are able to penetrate for
a skin depth inside the conductor. 

In our problem, we do not really have to take account of a change
of speed for charge carriers due to a finite resistivity: after 
all, magnetospheric electrons and positrons are likely to be always
fully relativistic, hence $1-v/c \ll 1$. Instead, we must take 
into account the fact that, as GWs drive the binary into ever 
closer orbits, the forcing-perturbation time-scale becomes comparable
to the time it takes perturbations to propagate inside the magnetosphere,
and thus to the time it takes to restore equipotentiality. This way 
the assumption that charge carriers are moving extremely fast is surely 
violated. 

We may derive the orbital time from
eq. \ref{eq:omega}, using $r=a/2$ as each star's distance from the
center of mass of the system, $T_{\rm orb} = 4\sqrt{2}\pi r^{3/2}/(GM)^{1/2}$.
For perturbations propagating in the poloidal plane, we may take
as the time to re-establish equilibrium the time $T_{\rm cc}$ that charge 
carriers take to follow (at speed $c$) a pure magnetic dipole field line;
these lines have equation $d = D\sin^2\theta$, where the  constant $D$ 
is the maximum distance of the line from the star, occurring on the 
equatorial plane $\theta = \pi/2$. Its length is
\begin{equation}
 L = \int_0^{\pi} r(\theta) d\theta = \frac{\pi}{2} D\;,
\end{equation}
hence $T_{\rm cc}(D) = \pi D/(2c)$. For the mid-distance line, $D = r$, we find:
\begin{eqnarray}
\chi & \equiv & \frac{T_{\rm cc}(D=r)}{T_{\rm orb}} = \frac{1}{8\sqrt{2}} \left(\frac{GM}{r c^2}\right)^{1/2} =  \nonumber \\ 
& = & 0.057 \left(\frac{M}{2.8\times M_\odot} \frac{10\text{ km}}{R_{\rm NS}} \frac{R_{\rm NS}}{r}\right)^{1/2}\;.
\end{eqnarray}
For perturbations propagating along the toroidal direction, we can directly 
compare the perturbation speed (which equals the orbital speed $v_{\rm orb}$) to the
speed of light, finding
\begin{eqnarray}
 \chi' = \frac{v_{\rm orb}}{c}  & = & 0.22 \left(\frac{GM}{r c^2}\right)^{1/2} =   \nonumber \\ 
 & = & 0.22 \left(\frac{M}{2.8\times M_\odot} \frac{10\text{ km}}{R_{\rm NS}} \frac{R_{\rm NS}}{r}\right)^{1/2}\;.
\end{eqnarray}
From these equations, we see that at large distances $\chi,\chi' \ll 1$, hence we expect 
an accurate screening to occur, but as the two stars approach, $\chi \approx 0.1, 
\chi'\approx 0.2$, which means the screening will be less than perfect. As a 
comparison, in a metallic onductor $\sigma \approx 5\times 10^{17} \; s^{-1}$, while, for optical
wavelengths, $\nu \approx 5\times 10^{14}\text{ Hz}$, giving $\chi = \nu/\sigma \approx 10^{-3}$. 

By analogy with the finite-resistivity conductor mentioned above, this implies that 
perturbation fields, due to the time-dependent perturbation caused by the companion
star, will penetrate for a finite length inside the (otherwise shielded) magnetosphere,
without the local charge carriers being able to short out the component of
$\vec E$ along $\vec B$, hence 
\begin{equation}
 \label{eq:unstransversality}
\vec E\cdot\vec B \neq 0\;.
\end{equation}
We can actually push the analogy further to discuss the skin depth.  In conducting media with 
conductivity $\sigma$ (in planar geometry!), the wavenumber is found to be \citep{Jackson75}:
\begin{equation}
 \label{eq:kmetal}
 k \approx  2\pi (1+\imath) \frac{\sqrt{\sigma\nu}}{c}
\end{equation}
where $\nu$ is the impinging wave frequency. In our case, we can take 
$\nu= 1/T_{\rm orb}$ and $\sigma \approx 1/T_{\rm cc}$, because $1/\sigma$ is the time 
scale over which a local charge excess spreads itself over the surface 
of a conductor, i.e., the time scale to restore equipotentiality, which is the
same physical interpretation of $T_{\rm cc}$. We obtain for the skin depth
\begin{equation}
 \label{eq:skindepth}
 \delta = c \left(T_{\rm cc} T_{\rm orb}\right)^{1/2}\;.
\end{equation}
As a sanity check, we check that for very large separations ($r\rightarrow\infty$)
shielding is found to be effective, as expected. In this limit, we cannot neglect the 
stellar rotation
rate $\Omega$ with respect to the orbital one, $\omega$, because 
$\Omega \gg \omega$; each star will then have its own corotating
magnetosphere out to a distance $c/\Omega$, with typical time
$T_{\rm cc}(c/\Omega) = \pi /2\Omega$ (or shorter for inner lines); in 
this case, $T_{\rm cc}$ does not depend on $r$, $T_{\rm orb} \propto r^{3/2}$, and the
ratio $\delta/r \propto r^{-1/4} \rightarrow 0$, showing that (at large 
separations) the skin depth is exactly that, a thin layer where 
perturbations can penetrate. It is worth remarking that, in the case of
the binary pulsar, $\delta/r \approx 4.8$ for $T_{\rm cc}(c/\Omega)$ for 
PSR J0737-3039A, and {\it a fortiori} $\delta/r > 1$ for the slower 
PSR J0737-3039B.

Conversely, when $r\rightarrow R_{\rm NS}$, 
\begin{equation}
 \frac{\delta}{r} = \frac{c T_{\rm cc}(r)}{r} \chi^{-1/2} = \frac{\pi}{2\chi^{1/2}} 
 = {\cal O}(1)
\end{equation}
which shows that the skin depth is as large as the whole magnetosphere. 
In this limit, the analysis leading 
to the expressions for $k$ and $\delta$ becomes invalid. In fact, objects of 
spherical symmetry smaller than about the mean free path for deflection of 
conduction electrons (i.e., in our case, $c T_{\rm cc}$) are generally 
considered to be only weakly affected by skin depth effects, in the sense that
the spherical metal conductor is not shielded from outside fields, see for 
instance \citet{Petrov81, Morokhov81}. For this reason, and because
$\delta \approx r$ at the moment of merger, we feel justified in neglecting 
skin depth effects in the latest stages of the binary evolution. 

Given the arbitrary orientation of the three axes involved, $\vec l, \vec \mu_1,
\vec \mu_2$, and the fact that the skin depth, in the late stages of the 
binary evolution, is of order of the orbital separation, $\delta \approx
r$, it seems difficult to escape the conclusion that, under these circumstances,
\begin{equation}
\label{eq:nontransversality}
\vec E \cdot \vec B \neq 0\;.
\end{equation}

\subsection{Numerical estimates}

We expect emission by particles accelerated by (unscreened) induction 
electric fields to be strongly concentrated toward the last moments 
before the binary merges, exactly like the quadrupole radiation we considered above. 
Hence we shall neglect shielding, as justified above. Also, we are 
interested in the total electromotive force along closed magnetic field lines, 
both because we still expect synchrotron losses to constrain 
particles to follow magnetic field lines, and because the corotating
magnetosphere is where the largest particle density is located, and
is thus likely to be where most radiation is emitted. 

In order to get an idea of the size of this effect, we use as a model 
the {\it in vacuo} magnetosphere, using once again 
Monhagan's expressions, since they represent the {\it exact} solutions 
for the fields of point--like magnetic dipole, in the {\it in vacuo} case. 
We choose arbitrary orientations of the magnetic dipoles such that 
$(\vec\mu_1\wedge\vec\mu_2)\cdot \vec l\neq 0$, where $\vec l$ is the 
orbital angular momentum. A configuration for the magnetic field for a
given, generic choice for the magnetic moments is given in fig. \ref{fig:field}.

\begin{figure}
\vspace{+0\baselineskip}
{
\includegraphics[width=0.45\textwidth]{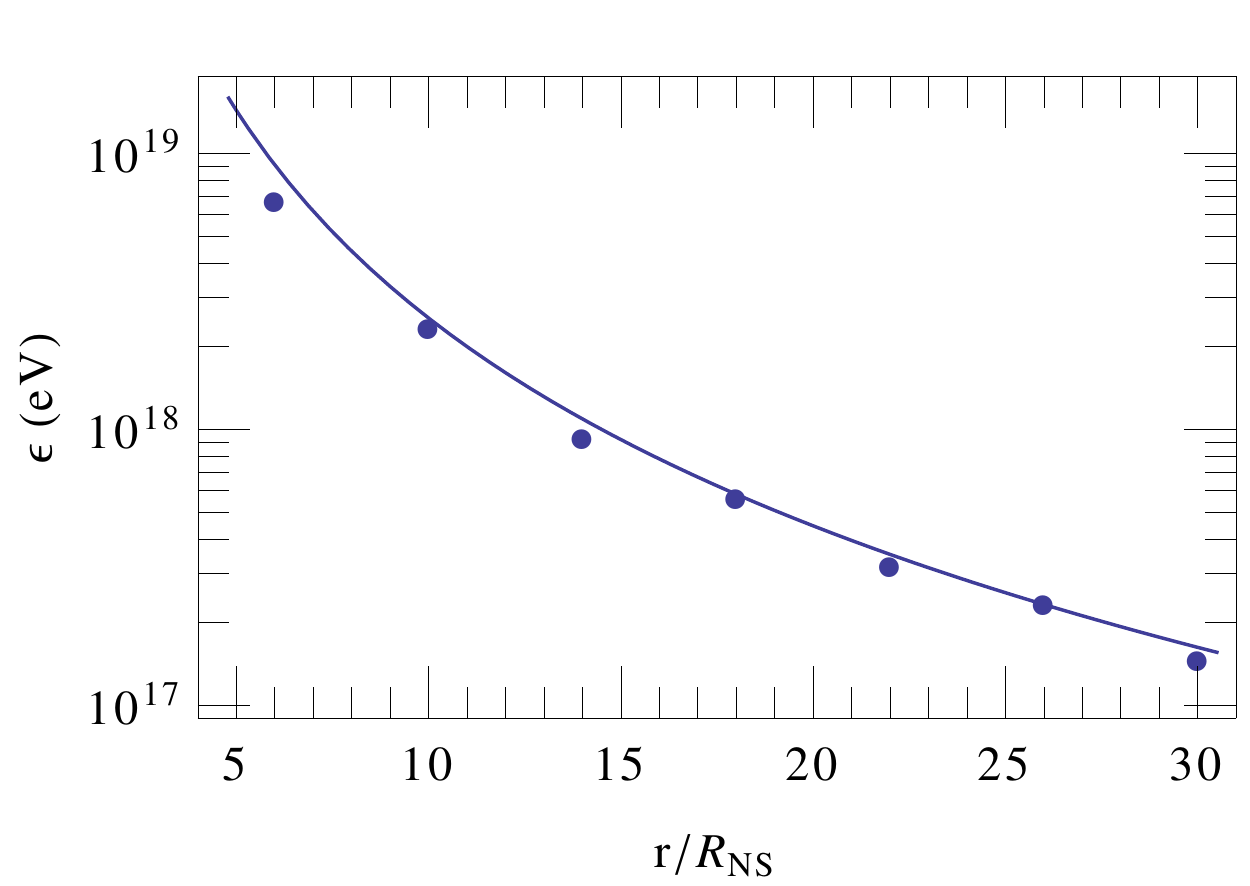}
}
\caption{The total electromotive force along a magnetic field line
closing at the midpoint between two stars of equal magnetic moments,
as the distance between the stars is reduced by GWR. The solid line is 
an $r^{-5/2}$ fit.
\label{fig:calE}
}
\vspace{-1\baselineskip}
\end{figure}

\begin{figure}
\vspace{+0\baselineskip}
{
\includegraphics[width=0.45\textwidth]{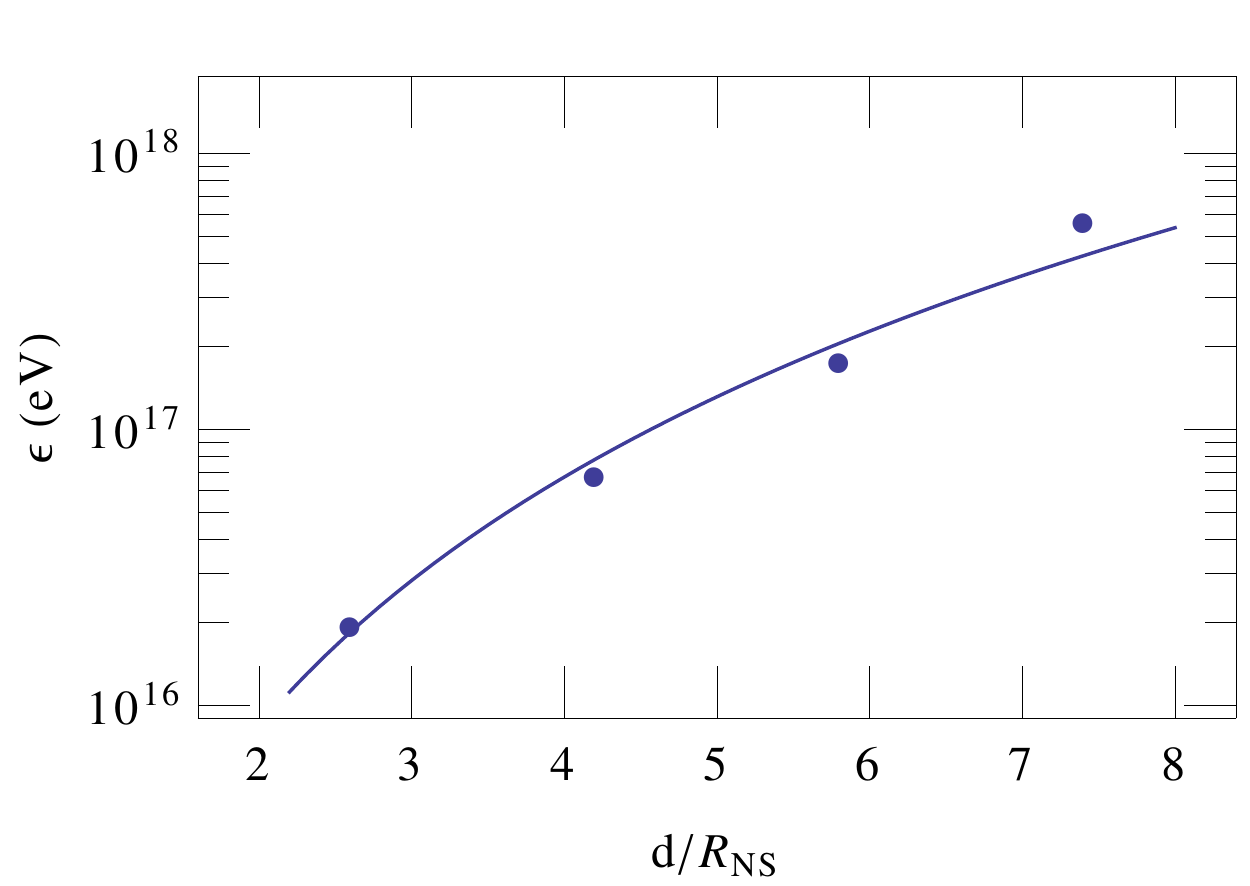}
}
\caption{The electromotive force for lines reaching only as far as $d$,
at a given instant. The solid line is a $d^3$ approximation.
\label{fig:innermost}
}
\vspace{-1\baselineskip}
\end{figure}

On the line connecting the two dipoles, the two magnetospheres 
touch approximately where $\mu_1/R_1^2 \approx \mu_2/R_2^3$, with 
$R_1+R_2$ the stellar separation. We now compute numerically, for the 
{\it in vacuo} case, the integral 
\begin{equation}
 {\cal E} = \int \vec E\cdot d\!\vec s
\end{equation}
for the part (outside the star) of a magnetic field line which 
passes close to $R_1$, i.e., roughly the line separating the two 
magnetospheres. The numerical result for the exact fields given
by Monhagan (eqs. \ref{eq:Eexact} and \ref{eq:Bexact}) is displayed 
in fig. \ref{fig:calE}. Here we have chosen $\mu_1 = \mu_2 = 10^{30}$ 
in cgs units, from which $R_1 \approx R_2$ (the approximate sign,
instead of the equality sign, results from the different inclinations of 
two dipoles). Please notice that the different dots refer to different 
field lines: we are comparing the electromotive force always for the
line separating the two magnetospheres, which changes as GW reduces
the stars' separation $2r$. 

Our approximation, ${\cal E} \propto r^{-5/2}$ is also displayed in fig. \ref{fig:calE}.
It can be obtained by means of the following argument. We are interested in 
${\cal E}$ when the stars are very close and in a region very close to {\it both}
stars, in which case we see from eqs. \ref{eq:Eexact} 
and \ref{eq:Bexact} that the fields are basically given by the formulas
for uniform translation: the remaining terms are corrections of order $\omega r/c$. 
Physically, since each star moves 
roughly in a straight path over a time scale shorter than $v/a$, where $a=v^2/r$ is
the acceleration, the fields inside a distance $c v/a$ are not affected by time-delay 
effects, and equal the fields due to a dipole in uniform translation. The 
condition $r < cv/a$ can easily be rewritten as
\begin{equation}
 r < \frac{c}{\omega}
\end{equation}
i.e., it restricts the domain of validity of the approximation to the volume 
inside the light cylinder defined in terms of the orbital angular speed, $\omega$. 
Then the electric field generated by star $1$ is
\begin{equation}
\vec{E}_1\approx\frac{\vec{v}_1}{c}\wedge\vec{B}_1\;,
\end{equation}
where $\vec{v}_1$ is the velocity of star $1$. This component of the electric 
field is perpendicular to the magnetic field $\vec{B}_1$ of the star itself.
We also add the contribution due to the {\it other} star (the one we called $2$) 
to the total electric field
\begin{equation}
\vec{E}=\vec{E}_1+\vec{E}_2\approx\frac{\vec{v}_1}{c}\wedge\vec{B}_1+\frac{\vec{
v}_2}{c}\wedge\vec{B}_2\;,
\end{equation}
which, in general, is not perpendicular to the total magnetic field 
$\vec{B}=\vec{B}_1+\vec{B}_2$. The component of the induction electric field 
along the magnetic field is
\begin{equation}
\label{eq:Eapprox}
E_\parallel=\vec{E}\cdot\frac{\vec{B}}{|\vec{B}|}\approx\frac{\vec{v}_1-\vec{v}_
2}{c}\cdot\frac{\vec{B}_1\wedge\vec{B}_2}{|\vec{B}|}=\frac{\vec{v}}{c}\cdot\left
(\hat{n}\wedge\vec{B}_2\right)\;,
\end{equation}
where $\vec{v}=\vec{v}_1-\vec{v}_2$ is the stars' relative speed and $\hat{n}$ 
is the unit vector along the magnetic field $\vec{B}$, in the case of nearly
equal--mass stars. Recalling that $B_2 \propto r^{-3}$, $v \propto r^{-1/2}$, and
the length over which $\vec E$ is to be integrated is $\propto r$, we arrive 
at the approximation ${\cal E} \propto r^{-5/2}$. 

The meaning of this approximation is the following.  Close to the point where 
the two magnetospheres have roughly equal values of $B_1\approx B_2\approx B$, 
we find $E_\parallel\approx vB_2/c\approx vB/c$, {\it i.e.} the parallel 
electric field is as big as the maximum achievable.
Far from this region, {\it i.e.} closer to the surface of star $1$ (where 
$B_2\ll B_1\approx B$), the magnetic line will be nearly exactly parallel to 
$\vec B_1$, and the induction field nearly due to the motion  star $1$ alone, 
$\vec E_1 \approx \vec{v}_1\wedge\vec B_1/c$: thus $E_\parallel\approx 
vB_2/c\ll vB/c$, {\it i.e.} the parallel electric field acts as a small 
perturbation.
 
As an order of magnitude, we take for $\vec E$ the  value estimated above for the 
{\it in vacuo} case:
\begin{equation}
 \label{eq:EM}
 {\cal E} \approx 7\times 10^{19} \eta \left(\frac{M}{2.8\times 
 M_\odot}\right)^{1/2} \left(\frac{B_2}{10^{12}\text{ G}}\right)
 \left(\frac{R_{\rm NS}}{r}\right)^{5/2}\text{ eV}\;,
\end{equation}
where $r=R/2$ is the half-distance between the stars, $B_2$ is the dipolar component of the magnetic 
field at the surface of star $2$, and $\eta\approx 1$ is a 
fudge factor depending on the relative directions of $\vec\mu_1$ and 
$\vec\mu_2$, the choice of the magnetic field line along which we perform the 
numerical integration, the choice of phase in its travel from dayside to 
nightside and {\it vice versa}\footnote{Strictly speaking, there are also 
relative orientations of $\vec\mu_1$ and $\vec\mu_2$ for which the above 
integral necessarily vanishes.
However, this happens only when both the magnetic moments are strictly parallel 
to $\vec{\omega}$, a sufficiently rare occurrence that we may neglect in the 
following.}. This analytical computation is borne out by a numerical 
integration, in fig. \ref{fig:calE}, for arbitrary orientation. The 
scaling with $r$ applies to all orientations we investigated. 

The reason why we chose, among all possible magnetic field line, the last 
one closing onto a star, in the numerical estimate above, is that 
${\cal E}$ has a maximum for this line. In all other cases, the total
magnetic field is dominated by the nearer star, the induction electric field 
is also dominated by the nearer star, but for these $\vec E_1\cdot \vec B_1 
= 0$. The only effect is due to the component $E_2 \approx v B_2/c$, which 
however decreases due to the larger distance from the star $2$. 
This is shown in fig. \ref{fig:innermost}, where ${\cal E}$ is computed 
for fixed orbital separation, but for different distances $d$ from star $1$
at which the line crosses the line joining the two stars' centers.

\section{Conclusions}

In this paper, we have begun an investigation of the electrodynamics of
binary pulsars. In analogy with IRAPs, we have first determined the rate of 
energy loss due to (quadrupolar, not dipolar) radiation, showing that it is 
dwarfed by GW losses, but may yet be detectable due its strong transient 
character, and its near periodicity. We have also discussed the presence of 
transient induction electric fields, which, {\it in vacuo}, cause 
extremely large electromotive forces along magnetic field lines in the 
corotating magnetosphere. We have also argued, on the basis of an analogy
with good conductors, that the skin-depth of these transient fields in a
realistic (i.e., charge-rich) magnetosphere, is large once the stars are close
to merging, while it is truly thin at large distances from the star. This 
implies that free charges in the joint magnetosphere will be subject to 
large electric fields with a component parallel to the total magnetic field.
The consequences of this simple observation will be discussed elsewhere.

\section*{Appendix}

Here we derive eq. \ref{eq:Ptotal} from eq. \ref{eq:Pint}. We take the axis
$z$ to be perpendicular to the plane of the binary star motion, so that, in 
Cartesian coordinates, $\vec{\ddot \beta} = \omega^2\beta (\cos\omega t, 
\sin\omega t, 0)$.  Also, $\vec\mu_1-\vec\mu_2 = (\mu_{\rm x}, \mu_{\rm y},\mu_{\rm z})$, and
$\hat n = (\cos\theta, \sin\theta\cos\phi, \sin\theta \sin\phi)$. We thus find
\begin{equation}
 \hat n\cdot(\vec\mu_1-\vec\mu_2) = \mu_{\rm x}\cos\theta + \mu_{\rm y}\sin\theta\cos\phi + \mu_{\rm z}
 \sin\theta\sin\phi\;,
\end{equation}
and
\begin{equation}
 (\vec{\ddot\beta}\wedge\hat n )^2 = \omega^4 \beta^2
 \left[1 - (\cos\omega t\cos\theta + \sin\omega t\sin\theta\cos\phi)^2\right]\;.
\end{equation}
The time average over the orbital period is now trivial, and we obtain:
\begin{eqnarray}
 P & = & \frac{\omega^4\beta^2}{4\pi c^3} \int^1_{-1} d\cos\theta \int_0^{2\pi} d\phi \times \nonumber \\
& & (\mu_{\rm x}\cos\theta + \mu_{\rm y}\sin\theta\cos\phi + \mu_{\rm z}\sin\theta\sin\phi)^2 \times \nonumber \\
& & (1-\frac{1}{2} \cos^2\theta - \frac{1}{2}\sin^2\theta\sin^2\phi) =  \nonumber \\
& = & \frac{\omega^4\beta^2 }{15 c^3}(3\mu_\perp^2 +4\mu_{\rm z}^2)
\end{eqnarray}

\bibliographystyle{mn2e}
\bibliography{ms}

\end{document}